\documentclass[12pt,preprint]{elsarticle}

\usepackage{bm,amsmath,amssymb}
\usepackage{graphicx}
\usepackage{subfigure}
\usepackage{color}
\usepackage{xcolor}
\usepackage{soul}
\usepackage{hyperref}
\usepackage{multirow}
\usepackage{lineno}
\usepackage{lipsum}
\usepackage{physics}
\hypersetup{
    colorlinks = false,
    urlcolor = red
    }

\newcommand \beqa {\begin{eqnarray}}
\newcommand \eeqa {\end{eqnarray}}

\usepackage{graphicx} 

\makeatletter
\def\ps@pprintTitle{%
 \let\@oddhead\@empty
 \let\@evenhead\@empty
 \def\@oddfoot{}%
 \let\@evenfoot\@oddfoot}
\makeatother

\begin{document}

\title{Lattice QCD at non-zero temperature and density}
\begin{frontmatter}

\author{F. Karsch}
\address{Fakult\"at f\"ur Physik, Universit\"at Bielefeld, D-33615 Bielefeld, Germany}

\begin{abstract}
Soon after the formulation of Quantum Chromodynamics in 1972
its regularization on Euclidean space-time lattices had been
introduced by Kenneth Wilson
\cite{Wilson:1974sk}. This 
paved ground for numerical studies of non-perturbative aspects of QCD, first shown by Michael Creutz
\cite{Creutz:1980zw}. Ever since
these first lattice QCD calculations the exploration of the 
QCD phase diagram and the thermodynamics of strong-interaction matter at
non-zero temperature and density
was pursued vigorously. In this 
brief review I try to highlight
some of the results on QCD thermodynamics obtain during
the last 42 years through lattice 
QCD calculations.

    


\end{abstract}
\end{frontmatter}

\section*{}
\vspace{-0.5cm}
   \begin{center}
    Article to appear in a special EPJC Volume in celebration of

’{\it 50 Years of Quantum Chromodynamics}’

edited by Franz Gross and Eberhard Klempt
\end{center}
\vspace{-0.5cm}
\noindent
\underline{\hspace{13.7cm}}

\section{QCD thermodynamics on Euclidean Lattices}
The path integral formulation of QCD can easily be applied to 
cases of non-vanishing temperature ($T$) and other
external control parameters, e.g. the chemical potentials ($\mu_f$) that couple to the conserved currents of quark-flavor number. 

Using the lattice regularization scheme of QCD, introduced by K. Wilson \cite{Wilson:1974sk},
QCD thermodynamics is formulated on Euclidean space-time
lattices of size $N_\sigma^3 N_\tau$
where, for a given lattice spacing ($a$), the lattice extent in Euclidean time controls
the temperature $T=1/N_\tau a$ and the spatial extent is related to the volume of
the thermodynamic system, $V=(N_\sigma a)^3$.
The chemical potentials enter directly in the fermion matrices, $M_f$, which arise from the
QCD Lagrangian after integrating out the
fermion fields. 

Bulk thermodynamics can then be derived from the lattice regularized partition function,
  \begin{eqnarray}
   Z = \int \prod_{x_0=1}^{N_\tau}
   \prod_{x_i=1}^{N_\sigma}
   \prod_{\hat{\nu=0}}^3\; \mathcal{D}U_{x,\hat{\nu}} \  e^{-S_G}
\prod_{f=u,d,s..}\det M_f(m_f,\mu_f) ~,
\label{partition}
  \end{eqnarray}  
where $x=(x_0,\vec{x})$ labels the sites
of the 4-dimensional lattice, $S_G$ denotes the gluonic part of the Euclidean action, which is expressed in terms of $SU(3)$ matrices $U_{x,\hat{\nu}}$
and $M_f$ is the fermion matrix for quark flavor $f$. It is a function of quark mass, $m_f$
and flavor chemical potential
$\hat{\mu}_f\equiv \mu_f/T$.
Basic bulk thermodynamic observables (equation of state, susceptibilities, etc.) can then be obtained from the logarithm of the partition function,
$Z$, which defines the pressure, $P$, as
\begin{equation}
    P/T = \frac{1}{V}\ln Z(T,V,\vec{\mu},\vec{m}) \; .
\end{equation}
Applying standard thermodynamic relations one obtains other observables of interest; e.g. the 
energy density is related to the 
trace anomaly of the energy-momen\-tum
tensor, $\Theta^{\mu\mu}$,
\begin{equation}
\frac{\Theta^{\mu\mu}}{T^4}
=\frac{\epsilon -3P}{T^4}
\equiv T\frac{\partial P/T^4}{\partial T}
    \;,
    \label{energy}
\end{equation}
and the conserved charge densities
are obtained as,
\begin{equation}
    \frac{n_X}{T^3} =\frac{\partial P/T^4}{\partial \hat{\mu}_X}\;\; ,\;\; X=B,\ Q,\ S \; .
\end{equation}

While the framework of lattice QCD
provides easy access to QCD thermodynamics at vanishing~ values of the chemical potentials, major 
difficulties arise at $\mu_f\ne 0$.
The fermion determinants,
${\rm det}M_f(m_f,\mu_f)$, are no
longer positive definite
when the real part of the chemical potential is non-zero, 
${\rm Re} \hat{\mu}_f\ne 0$. This includes the physically relevant case of strictly real chemical potentials. The presence of a complex valued integrand in the path integral makes the 
application of standard Monte Carlo techniques, which rely on a probabilistic interpretation of integration measures, impossible.
The two most common approaches  
to circumvent this problem
are to either (i) perform numerical calculations at imaginary values of the chemical potential, $\hat{\mu}_f^2<0$ \cite{DElia:2002tig,deForcrand:2002hgr},
or to (ii) perform Taylor series expansions 
around $\hat{\mu}_f=0$ \cite{Gavai:2001fr,Allton:2002zi}. 
In the former case numerical results need to be analytically 
continued to real values of $\mu_f$.
In the latter case
the QCD partition function is written as,
\begin{equation}
P/T^4 = \frac{1}{VT^3}\ln Z(T,V,\vec{\mu}) 
= \sum_{i,j,k=0}^\infty 
\frac{\chi_{ijk}^{BQS}}{i!j!k!}
\hat{\mu}_B^i \hat{\mu}_Q^j \hat{\mu}_S^k  \; ,
\label{Pdefinition}
\end{equation}
with $\chi_{000}^{BQS}\equiv P(T,V,\vec{0})/T^4$
and expansion coefficients,
\begin{equation} 
\chi_{ijk}^{BQS}(T) =\left. 
\frac{\partial P/T^4}{\partial\hat{\mu}_B^i \partial\hat{\mu}_Q^j \partial\hat{\mu}_S^k}\right|_{\hat{\mu}=0} \; ,
\label{suscept}
\end{equation}
can be determined in Monte Carlo simulations performed at $\hat{\mu}_X=0$.

The phase structure of QCD can be 
explored using suitable observables
that are sensitive to the spontaneous breaking and the eventual restoration of global symmetries. They can
act as order parameters in certain limits of the parameter space spanned by the quark masses. 
In QCD exact symmetries exist either in the chiral limit, {\it i.e.} at vanishing values of $n_f$ quark masses, or for 
infinitely heavy quarks, {\it i.e.}
in pure $SU(N_c)$ gauge theories, with
$N_c$ denoting the number of colors.

In order to probe the restoration of the global chiral symmetries one
analyzes the chiral condensate and
its susceptibilities,
\begin{eqnarray}
\langle \bar{\chi}\chi\rangle_f &=&
   \frac{T}{V} \frac{\partial}{\partial m_f} \ln Z
    = \frac{T}{V}  \langle{{\rm Tr} M_f^{-1}}\rangle \; ,
\label{condesate} \\
\chi_m^{fg} &=&\frac{\partial \langle \bar{\chi}\chi\rangle_f}{\partial m_g}
\; ,\; \quad
\chi_t^{f} =T \frac{\partial \langle \bar{\chi}\chi\rangle_f}{\partial T} \; .
\label{chiral-sus}
\end{eqnarray}
The former is an order parameter for 
the restoration of the $SU(n_f)_L\times SU(n_f)_R$ chiral flavor symmetry of QCD and
distinguishes, in the limit of vanishing
quark masses, a symmetry broken phase at 
low temperature from a chiral 
symmetry restored phase at high temperature,
\begin{equation}
\lim_{m_\ell\rightarrow 0} 
\langle \bar\chi \chi\rangle_\ell 
\begin{cases}
> 0 &,\;\; T< T_\chi \\
= 0 &\;, \;\; T \ge T_\chi
\end{cases}
 \; .
 \label{order}
\end{equation}
Similarly one considers 
the Polyakov loop $\langle L\rangle$ and its susceptibility $\chi_L$,

\begin{eqnarray}
\langle L\rangle &=& 
\frac{1}{N_\sigma^3}\langle \sum_{\vec{x}} {\rm Tr}L_{\vec{x}} \rangle \;\; ,
\;\; \quad L_{\vec{x}} = \prod_{x_0=1}^{N_\tau} U_{(x_0,\vec{x}),\hat{0}} \; ,
\nonumber \\
\chi_L &=& N_\sigma^3 \left(\langle L^2\rangle - \langle L\rangle^2 \right) \; ,
\label{Polyakov}
\end{eqnarray}
to probe the breaking and restoration
of the global $Z(N_c)$ center symmetry of 
pure $SU(N_c)$ gauge theories; {\it i.e.}
$SU(N_c)$ gauge theories at finite temperature, formulated on Euclidean lattices, are invariant under global
rotation of all temporal gauge field variables, $U_{\vec{x},\hat{0}}\ \rightarrow\ z U_{\vec{x},\hat{0}}$, with $z\in Z(N_c)$. The Polyakov loop expectation
value vanishes as long as this 
center symmetry is not spontaneously
broken.

The Polyakov loop expectation value also
reflects the long distance behavior 
of Polyakov loop correlation functions,
\begin{eqnarray}
\left| \langle L\rangle\right|^2 &\equiv&
\lim_{|\vec{x}|\rightarrow\infty}
G_L(\vec{x}) 
\begin{cases}
= 0  \Leftrightarrow F_q = \infty\; , &\hspace{-0.3cm} T\le T_{d} \\
> 0 \Leftrightarrow F_q < \infty\; , &\hspace{-0.3cm} T> T_{d}
\end{cases}
\end{eqnarray}
where 
\begin{eqnarray}
G_L(\vec{x}) &=& {\rm e}^{-F_{\bar{q}q}(\vec{x},T)} =
\langle {\rm Tr} L_{\vec{0}} {\rm Tr} L^\dagger_{\vec{0}} \rangle 
\end{eqnarray}
is the correlation function of two Polyakov loops. It denotes the change
in free energy (excess free energy, $F_{\bar{q}q}$),
that is due to the presence of two
static quark sources introduced in a thermal medium. At
zero temperature this free energy reduces to the potential between
static quark sources.

At least in the case of pure gauge theories this provides a connection
between
the confinement-deconfine\-ment phase transition and the breaking of a global 
symmetry, the $Z(N_c)$ center symmetry
of the $SU(N_c)$ gauge group. This symmetry, however, is 
explicitly broken in the presence of
dynamical quarks with mass $m_f<\infty$. Unlike chiral symmetry
restoration, deconfinement thus is 
not expected to be related to a 
phase transition in QCD with physical quark masses. Nonetheless,
the consequences of deconfinement,
related to the dissolution of hadronic bound states, 
becomes clearly visible in many thermodynamic observables.

\begin{figure*}[t]
    \centering
    \includegraphics[width=0.95\textwidth]{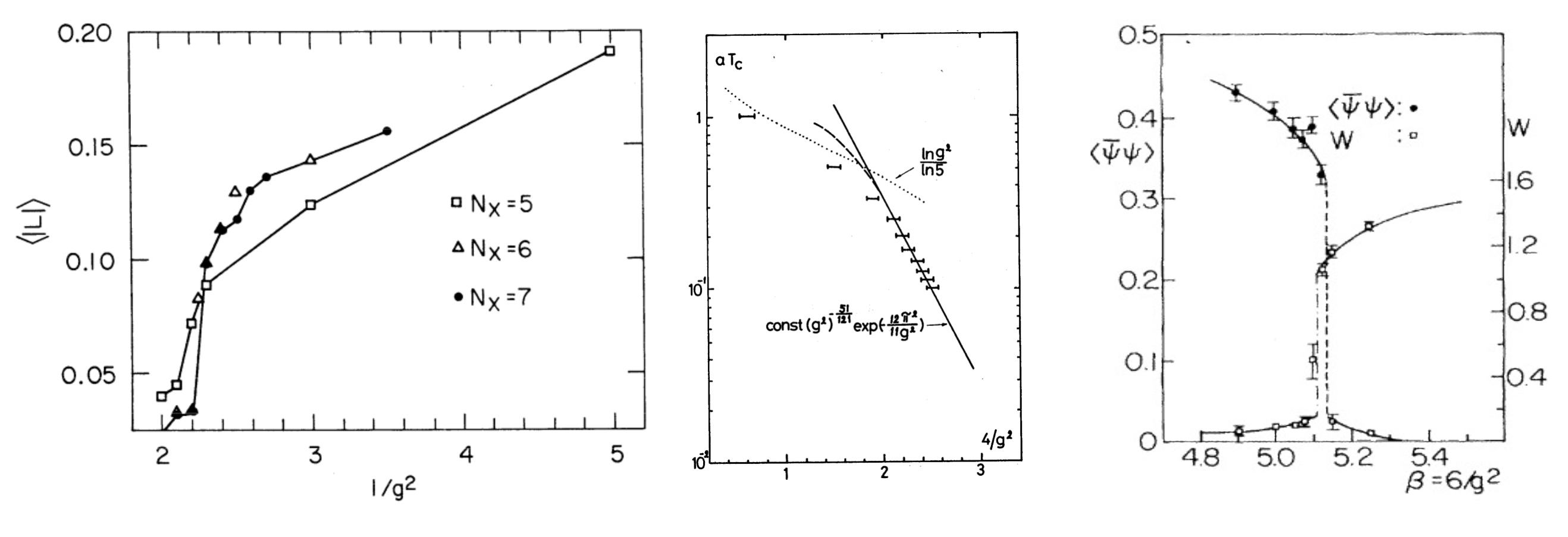}
    \caption{First evidence for the
    existence of a deconfinement phase transition in $SU(2)$ gauge theories using the Polyakov loop expectation value as an order parameter (left) \cite{McLerran:1980pk}  and a
    first extrapolation of the phase transition temperature to the continuum limit (middle) \cite{Kuti:1980gh} . The right hand figure shows a first
    comparison of the temperature dependence of the Polyakov loop ($W\equiv \langle |L|\rangle$) and chiral condensate 
    ($\langle \bar{\psi}\psi\rangle$) order parameters in a SU(3) gauge theory \cite{Kogut:1982rt} .
}
    \label{fig:1981}
\end{figure*}

\section{Early lattice QCD calculations at non-zero temperature}

Almost immediately after the formulation of QCD as the 
theory of strong interaction physics, its consequences 
for strong interaction matter at non-zero temperature were examined 
\cite{Collins:1974ky,Cabibbo:1975ig}.
It rapidly became obvious that fundamental properties of
QCD, confinement and asymptotic freedom on the one hand 
\cite{Cabibbo:1975ig,Baym:1979etb},
and chiral symmetry breaking on the other hand \cite{Pisarski:1983ms}, are likely to trigger a phase transition
in strong interaction matter that separates a phase being 
dominated by hadrons as the relevant degrees of freedom from
that of almost free quarks and gluons. The notion of a 
quark-gluon plasma was coined at that time \cite{Shuryak:1978ij}.

Soon after these early, conceptually important developments it was realized 
that the formulation of QCD on discrete space-time lattices, which
was introduced by K. Wilson as a regularization scheme in QCD 
\cite{Wilson:1974sk}, also provides a powerful framework for the analysis of non-perturbative
properties of strong interaction matter through Monte-Carlo simulations \cite{Creutz:1980zw}. 
This led
to a first determination of a phase transition temperature in $SU(2)$  
\cite{McLerran:1980pk,Kuti:1980gh} 
and $SU(3)$ \cite{Kogut:1982rt,Kajantie:1981wh,Yaffe:1982qf} gauge theories, and a first 
determination of the equation of state of purely
gluonic matter \cite{Engels:1980ty,Engels:1981qx}.
The interplay between deconfinement
on the one hand and chiral symmetry
restoration on the other hand also was studied \cite{Kogut:1982rt}
early on and the question whether or
not these two aspects of QCD may 
lead to two distinct phase transitions
in QCD has been considered ever since.
Some results from these first lattice 
QCD studies of the thermodynamics of strong interaction matter are shown in Fig.~\ref{fig:1981}. 

\begin{figure}[t]
\begin{minipage}[t]{0.47\textwidth}
\begin{center}
        \includegraphics[width=1.00\textwidth]{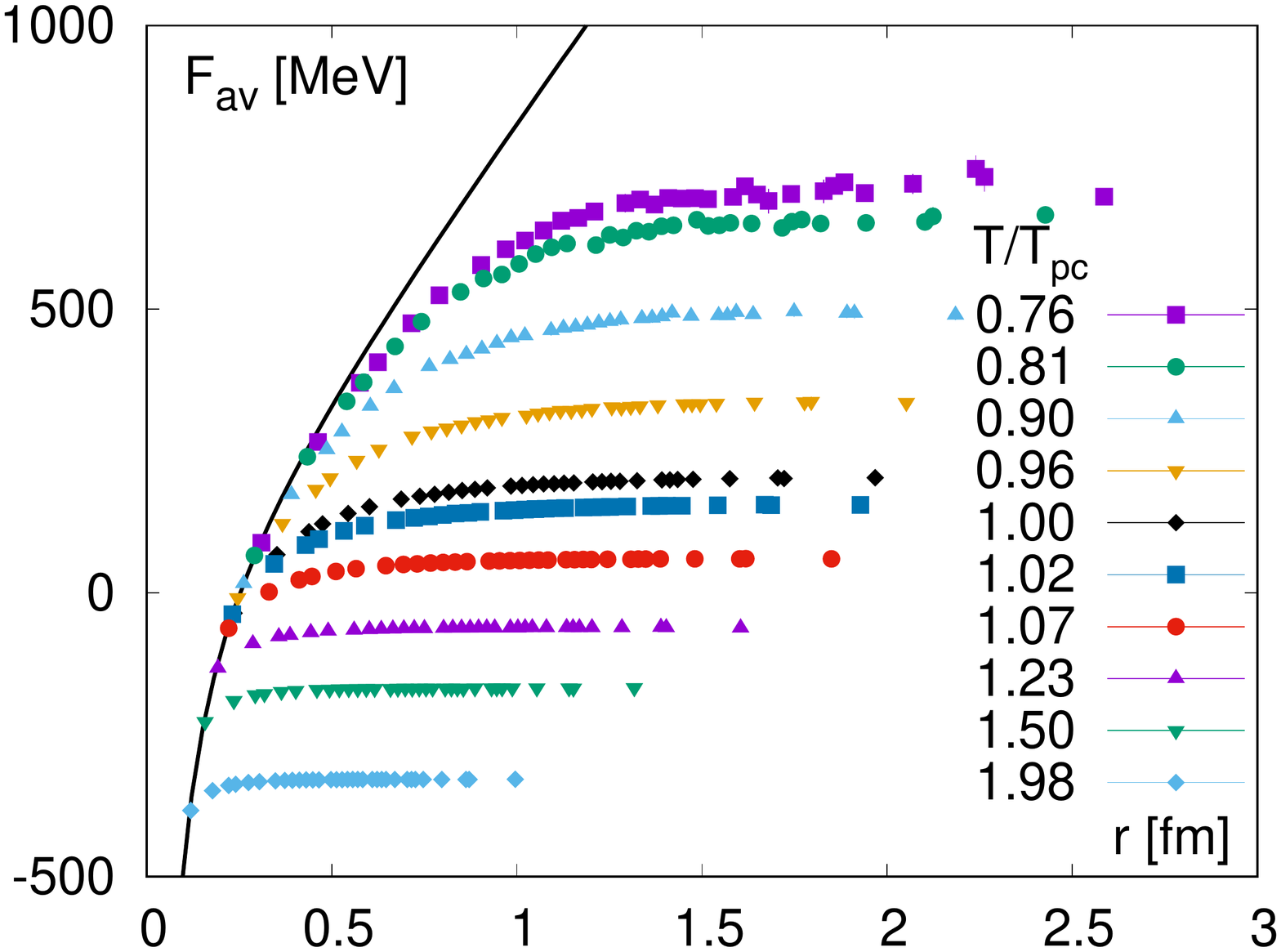}
        \end{center}
        \vspace*{-0.4cm}
\caption{The so-called color averaged, heavy quark free energy ($F_{\rm av}\equiv F_{\bar{q}q}$) in the vicinity of the pseudo-critical transition temperature ($T_{pc}$) in 2-flavor QCD \cite{Kaczmarek:2005ui}. Results shown cover a temperature range from $T/T_{pc}\simeq 0.75$ to $T/T_{pc}\simeq 2$.}
\label{fig:fea}
\end{minipage}
\begin{minipage}[t]{0.01\textwidth}
~
\end{minipage}
\begin{minipage}[t]{0.49\textwidth}
\centering
\includegraphics[width=0.93\textwidth]{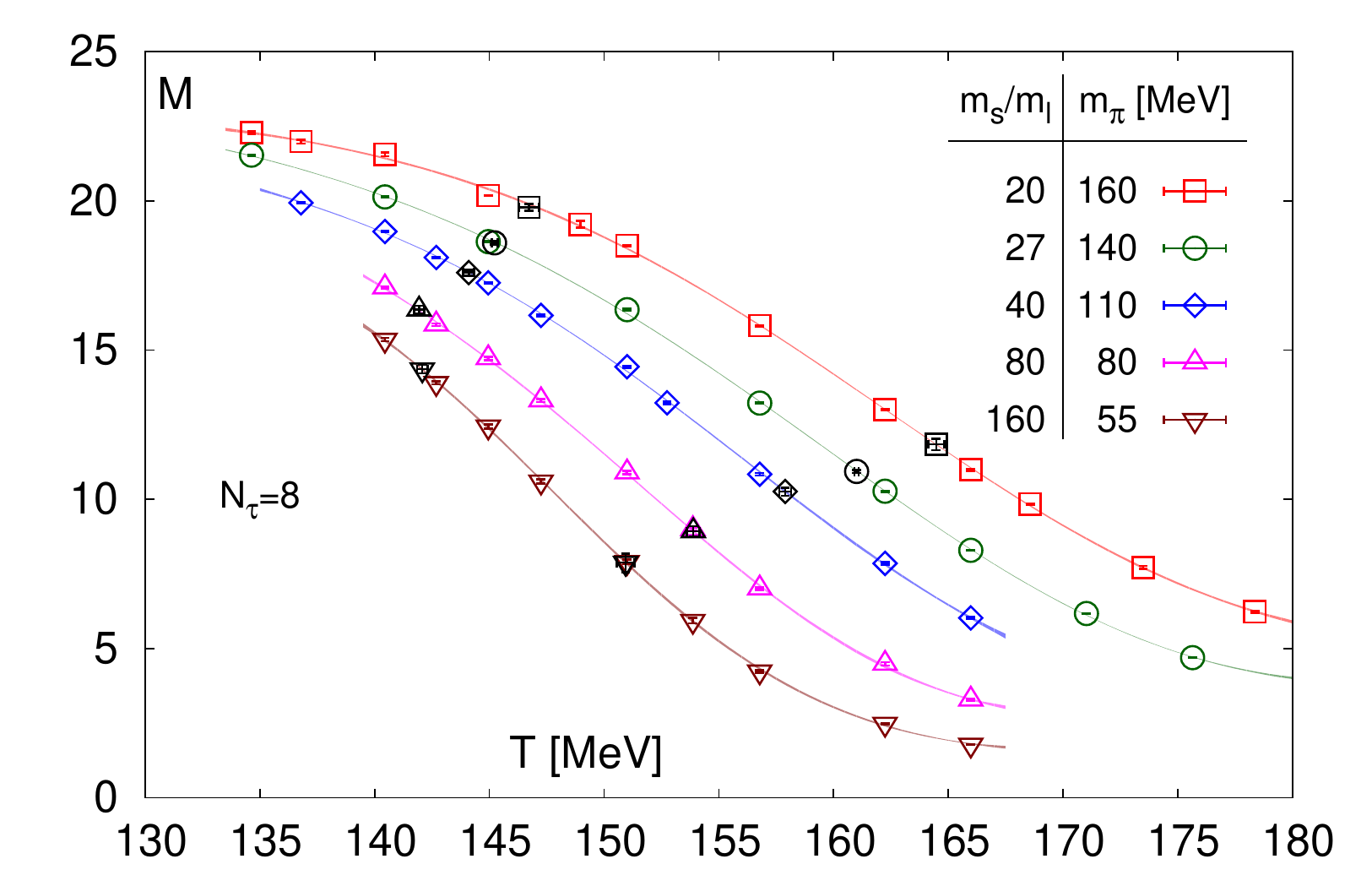} ~~

\caption{Quark mass dependence of chiral order parameter, $M$, defined in Eq.~\ref{Mren} for QCD with two degenerate light quark masses and a strange quark mass tuned to its physical value. Shown are results from calculations on lattices with temporal extent $N_\tau=8$ performed for several values of the light quark masses
\cite{HotQCD:2019xnw,Kaczmarek:2020err}. The light quark masses, $m_\ell$,
are expressed in units of the strange quark mass, $H=m_\ell/m_s$. In the figure we give $1/H = m_s/m_\ell$ together with the corresponding values of the Goldstone pion mass.
}
\label{fig:condensate}
\end{minipage}
\end{figure}

At physical values of the quark masses,
neither deconfinement nor the effective restoration of chiral
symmetry leads to a true phase transition. Still the transition from
the low temperature hadronic to the high temperature partonic phase of QCD
is clearly visible in the pseudo-critical behavior of the 
heavy quark free energy and the chiral
condensate respectively. Some recent
results on these observables, obtained in simulations of QCD with
light, dynamical quark degrees of freedom, are shown in
Figs.~\ref{fig:fea} and \ref{fig:condensate}.

\section{Global symmetries and the QCD phase diagram}

The early studies of QCD thermodynamics made it clear that universality arguments and 
renormalization group techniques, 
successfully developed in condensed matter physics and applied in statistical physics to the 
analysis of phase transitions, also
can be carried over to the analysis
of the phase structure of quantum field theories \cite{Rajagopal:1992qz,Rajagopal:2000wf}. The renormalization
group based arguments for the 
existence of a second order phase
transition in the universality class of the 3-d Ising model in a $SU(2)$ gauge theory, and a first
order transition for the $SU(3)$ color group of QCD \cite{Svetitsky:1982gs}
have been confirmed by detailed lattice QCD calculations \cite{Brown:1988qe,Engels:1989fz}.

In the presence of $n_f$ light, dynamical quarks, distinguished by a flavor quantum number, it is the chiral symmetry of QCD that triggers the occurrence of phase transitions \cite{Pisarski:1983ms}.
In addition to a global $U(1)$ symmetry that reflects the 
conservation of baryon number and 
is unbroken
at all temperatures and densities,
the massless QCD Lagrangian
is invariant under the
symmetry group
  \begin{equation}
   U(1)_A\times SU(n_f)_L \times SU(n_f)_R \;. 
   \label{Eq:symgroup}
  \end{equation}
The $SU(n_f)_L \times SU(n_f)_R$
symmetry
corresponds to chiral rotations of
$n_f$ massless quark fields in flavor space. This symmetry
is spontaneously broken at low 
temperatures, giving rise to $n_f^2-1$
massless Goldstone modes, which for
$n_f=2$ are the three light pions
of QCD. They have a non-vanishing mass only because of the 
explicit breaking of chiral
symmetry by a mass term in the QCD
Lagrangian that couples to the 
chiral order parameter field
$\bar{\chi}_f\chi_f$. 
The axial $U(1)_A$ group corresponds to global rotations of quark fields
for a given flavor $f$. Although it is an exact symmetry of the classical
Lagrangian, it is explicitly broken 
in the quantized theory. This explicit 
breaking of a global symmetry, 
arising from fluctuations on the
quantum level, is known as the $U(1)_A$ anomaly.

The renormalization group based
analysis of the chiral phase transition, performed by Pisarski and Wilczek \cite{Pisarski:1983ms}, made
it clear that the chiral phase transition is sensitive to the number 
of light quark flavors that become massless. Furthermore, it has been argued in \cite{Pisarski:1983ms}
that the order of the transition may be sensitive to the magnitude of the axial anomaly at non-zero temperature, 
which is closely related to the 
temperature dependence of topological 
non-trivial field configurations.

Although it was generally expected that the chiral phase transition in
3-flavor QCD becomes a first order phase transition in the chiral limit
\cite{Pisarski:1983ms}, there is currently
no direct evidence for this from lattice QCD calculations. In fact,
the current understanding is that the chiral phase
transition is second order for all $n_f\le 6$
\cite{Cuteri:2021ikv}. 

\begin{figure}[t]
\begin{center}
        \includegraphics[width=0.46\textwidth]{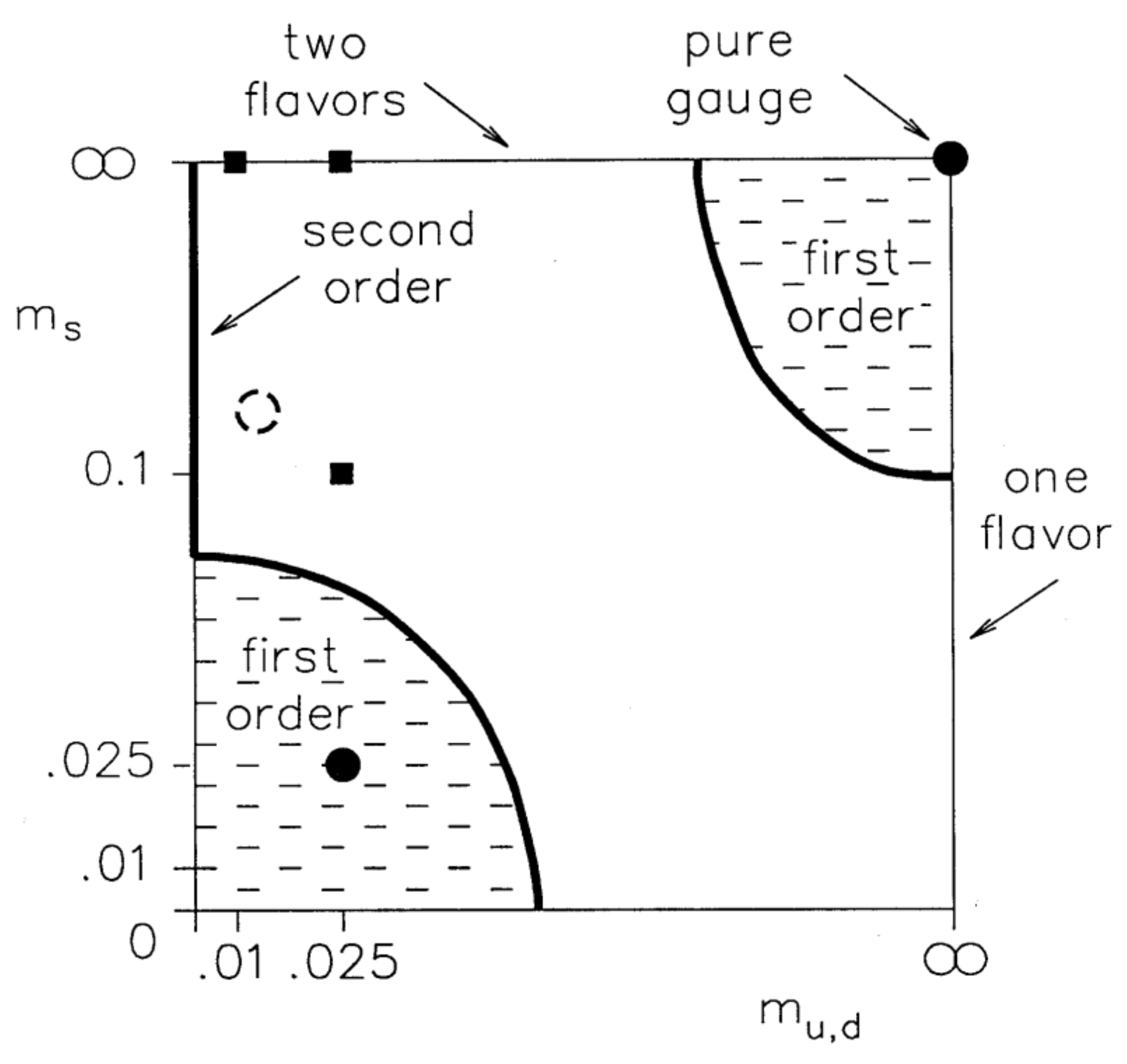}
        \includegraphics[width=0.42\textwidth]{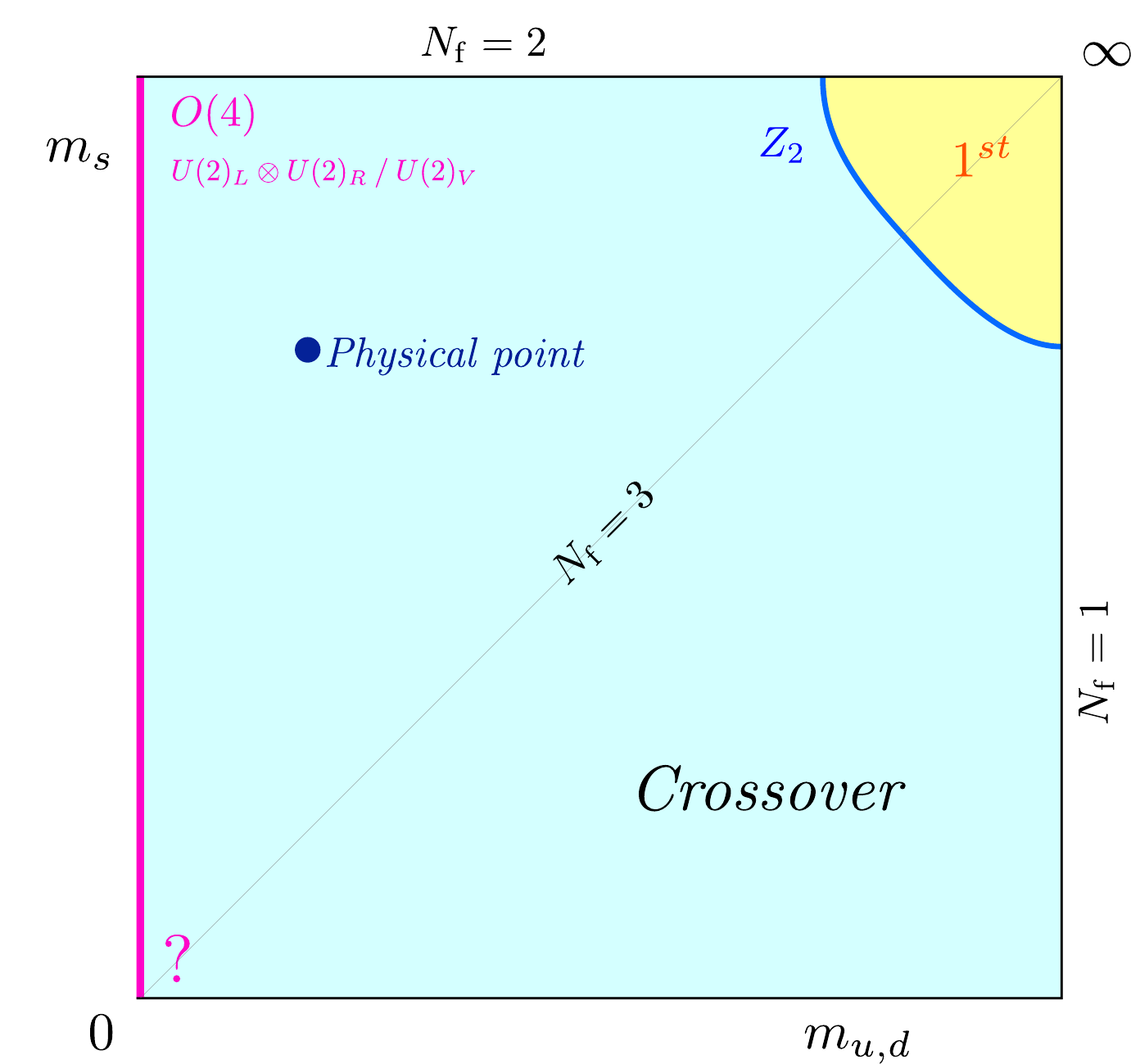}
        \end{center}
\caption{Sketch of the phase diagram of QCD  in the plane of degenerate, light up and down quark masses and a strange quark mass
(Columbia plot). The
figure shows the original version from 1990 \cite{Brown:1990ev} (left) and an updated version from 2021 \cite{Cuteri:2021ikv} (right). 
        }
\label{fig:Columbia}
\end{figure}

In Fig.~\ref{fig:Columbia}~(left) we show the original
version of the QCD phase diagram in the plane of two degenerate light $(m_\ell)$ and 
strange ($m_s$) quark masses, proposed in 1990 \cite{Brown:1990ev}, together
with an updated version from 2021 \cite{Cuteri:2021ikv}. Here $m_\ell$
denotes the two degenerate up and down
quark masses, $m_\ell\equiv m_u=m_d$.
This sketch of our current understanding of the 3-flavor phase diagram also is supported by the
increasing evidence for a non-singular crossover transition
in QCD with physical light and strange quark masses and the absence of any evidence for a first order
phase transition at lighter-than-physical values of 
the light and strange quark masses
\cite{Cuteri:2021ikv,Dini:2021hug}. 
In the chiral limit,
{\it i.e.} for vanishing
up and down quark masses\footnote{Lattice QCD studies 
of the (2+1)-flavor phase diagram generally are performed with
degenerate up and down quark masses.},
a second order phase transition 
will then occur.

\section{The chiral phase transition at vanishing chemical potential}

The occurrence of the chiral phase
transition is signaled
by the vanishing of the light 
quark chiral condensate. In order
to remove multiplicative and additive divergences in 
$\langle \bar\chi \chi\rangle_\ell$
one considers instead the
order parameter $M$ which is a 
combination of light and strange 
quark condensates,
\begin{eqnarray}
M &=& 2 \left( m_s \langle \bar\psi \psi\rangle_\ell - m_\ell \langle \bar\psi \psi\rangle_s
\right)/f_K^4 \; ,
\label{Mren}
\end{eqnarray}
and its derivative with respect to the light
quark masses, i.e. the chiral susceptibility $\chi_M$
\begin{eqnarray}
    \chi_M &=& m_s 
\left( 
\frac{\partial M}{\partial m_u} 
+\frac{\partial M}{\partial m_d}
\right)_{m_u=m_d\equiv m_\ell} \; .
\label{chiM}
\end{eqnarray}
Here the kaon decay constant
$f_K=156.1(9)/\sqrt{2}$~MeV, has 
been used to introduce a dimensionless order parameter.
The scaling behavior of $M$ and $\chi_M$, have 
been used to characterize the
chiral phase transition,
\begin{eqnarray}
M~~~~
&{\raise0.6ex\hbox{$\sim$\kern-0.75em\raise-1.4ex\hbox{\hspace*{-9pt}$m_\ell \to 0$}}}&~~
\begin{cases}
A \left( \frac{T_c^0-T}{T_c^0}\right)^{\beta}
& ,\;\; T< T_c^0 \\
~~0 & , \;\; T\ge T_c^0
\end{cases}
\label{M} 
\end{eqnarray}
\begin{eqnarray}
\chi_M~~~
&{\raise0.6ex\hbox{$\sim$\kern-0.75em\raise-1.4ex\hbox{\hspace*{-9pt}$m_\ell \to 0$}}}&~~
\begin{cases}
~~\infty & ,\;\; T\le T_c^0 \\
C \left( \frac{T-T_c^0}{T_c^0}\right)^{-\gamma}
& ,\;\; T> T_c^0 
\end{cases} \;
\end{eqnarray}
where $\beta,\ \gamma$ are critical 
exponents. 

We note that
the low temperature behavior of the 
order parameter susceptibility, $\chi_M$, is quite different from that
known, for instance, in the $3$-$d$ Ising model. The susceptibility diverges in the massless limit 
at all values of the temperature, 
$T \le T_c^0$. This is a consequence of the breaking of a continuous rather than a discrete symmetry. The former 
gives rise to Goldstone modes, the pions in QCD, which contribute to
the chiral condensate and as such to
the order parameter $M$, {\it i.e.},
\begin{equation}
    M \sim a(T)\sqrt{m_\ell}\;\; , \;\; T<T_c^0 \; .
\end{equation}
As a consequence the chiral susceptibility diverges below $T_c^0$,
$\chi_M \sim 1/\sqrt{m_\ell}$, while
at $T_c^0$ its divergence is controlled by the critical exponent
$\delta=1+\gamma/\beta$,
\begin{equation}
    \chi_M \sim
    \begin{cases}
    H^{-1/2} & \; ,\;  T< T_\chi \\
    H^{1/\delta -1} &\; ,\;  T=T_\chi
    \end{cases}
    \; ,
\end{equation}
with $H= m_\ell / m_s$. As $1-1/\delta >1/2$ in all relevant universality classes $\chi_M$ develops 
a pronounced peak at small, but
non-zero values of the quark masses,
\begin{equation}
    \chi_M^{peak}\equiv \chi_M(T_{pc}(H)) \sim H^{1/\delta-1}\; ,\; H= m_\ell / m_s\; .
\end{equation}
The location of such a peak in either
$\chi_M$ or similarly in $T\partial M/\partial T$, defines pseudo-critical
temperatures, $T_{pc}(H)$, which converge to the unique chiral
phase transition, $T_c^0$, at $H=0$.
\begin{figure}[!t]
\centering
\includegraphics[width=0.60\textwidth]{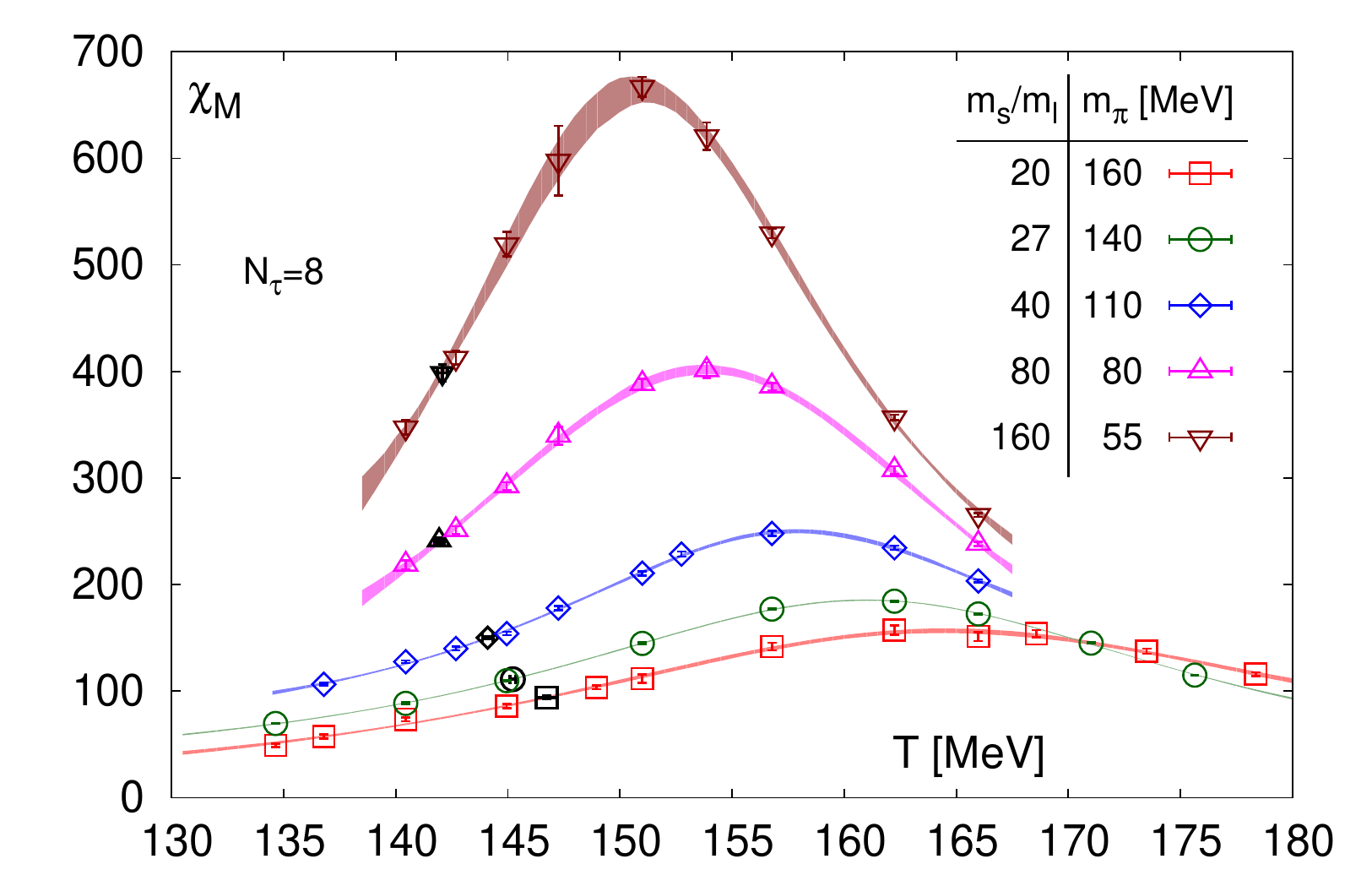}
\caption{ same as Fig.~\ref{fig:condensate} but for
the chiral susceptibility.
}
\label{fig:massvar}
\end{figure}
Some results on the quark mass dependence of $M$ and $\chi_M$ are
shown in Figs.~\ref{fig:condensate} and \ref{fig:massvar}, respectively.
A scaling analysis of these  
observables, performed in 
\cite{HotQCD:2019xnw}, led to the 
determination of the chiral phase transition temperature \cite{HotQCD:2019xnw},
\begin{equation}
    T_c^0=132^{+3}_{-6}\ {\rm MeV} \; .
\end{equation}
Similar results have also been obtained in \cite{Kotov:2021rah} where a quite different discretization scheme for the fermion sector of QCD has been used.

For physical light and strange quark masses, corresponding to $H\simeq 1/27$, one finds as a pseudo-critical
temperature \cite{HotQCD:2018pds},
\begin{equation}
    T_{pc}=156.5(1.5)\ {\rm MeV} \; ,
\end{equation}
which is in good agreement with
other determinations of pseudo-critical temperatures in
$(2+1)$-flavor QCD
\cite{Bonati:2015bha,Bonati:2018nut,Borsanyi:2020fev}.

The chiral symmetry group
$SU(2)_L\times SU(2)_R$ is 
isomorphic to
the rotation group $O(4)$. It thus is expected that the chiral phase transition for two vanishing light 
quark masses is in the same universality class as $3$-$d$, $O(4)$ symmetric spin
models. In fact, the rapid rise of 
$\chi_M$, shown in Fig.~\ref{fig:massvar}, is consistent with a critical exponent in this universality class, $\delta=4.824$ 
\cite{Guida:1998bx}. However, a precise
determination of this exponent in 
2-flavor QCD is not yet possible.
This leaves
open the possibility for other symmetry breaking 
patterns and other 
universality classes playing a role in the chiral limit of 2-flavor QCD
\cite{Pelissetto:2013hqa}. In fact,
the discussion of such possibilities
is closely related to the yet 
unsettled question
concerning the influence of the axial $U(1)_A$ symmetry on the chiral phase transition. For a recent review on
this question see, for instance 
\cite{Lahiri:2021lrk}.

\noindent
\underline{\it Thermal masses and screening masses:}
The restoration of symmetries is reflected also in the modification of the 
hadron spectrum at non-zero temperature. Interactions in a thermal medium lead
to modifications of resonance peaks that can modify the location of maxima and the 
width of spectral functions that control properties of hadron correlation functions.
This gives rise to so-called thermal masses as well as thermal screening masses that
control the long-distance behavior of hadron correlation functions in Euclidean time and spatial 
directions, respectively.

A consequence of $U(1)_A$ breaking in the vacuum or at low temperature
is that masses of hadronic states that are related to each other through a $U(1)_A$ transformation differ, while 
they become identical, or close to each other, when the $U(1)_A$ symmetry is 
effectively restored. This is easily
seen to happen at high temperature.
The crucial question, of relevance for the QCD phase transition, however, is 
to which extent $U(1)_A$ symmetry breaking is reduced, or already disappeared at the chiral phase transition temperature. Settling this question requires the analysis of
observables sensitive to $U(1)_A$
breaking close to $T_c^0$ and for
smaller-than-physical light
quark masses.

\begin{figure}[!t]
\centering
\includegraphics[width=0.54\textwidth]{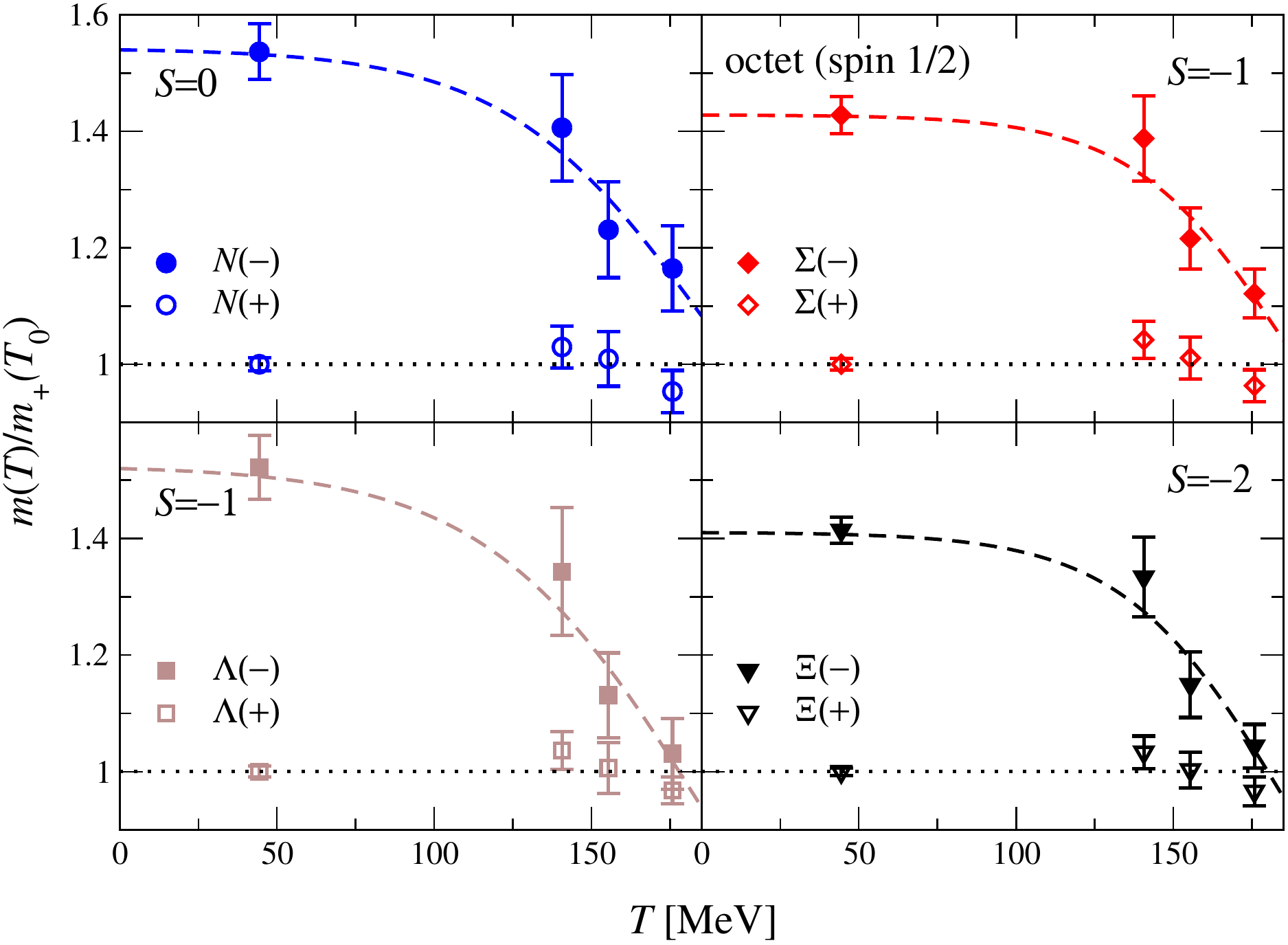}
\caption{Temperature dependence of 
masses of parity partners in the 
baryon octet
\cite{Aarts:2018glk}.
}
\label{fig:Tmass}
\end{figure}

The calculation of in-medium
modifications of hadron masses is
difficult, but has been attempted for quark masses close
to their physical values \cite{Aarts:2018glk}. Results for
the temperature dependence of the mass-splitting of parity partners in the baryon octet \cite{Aarts:2018glk}
are shown in Fig.~\ref{fig:Tmass}. These 
results suggest a strong temperature
dependence of the negative parity states while the 
positive parity partners are not sensitive to 
temperature changes. At $T_{pc}$
the masses of parity partners are 
almost degenerate.

\begin{figure}[!t]
\centering
\includegraphics[width=0.38\textwidth]{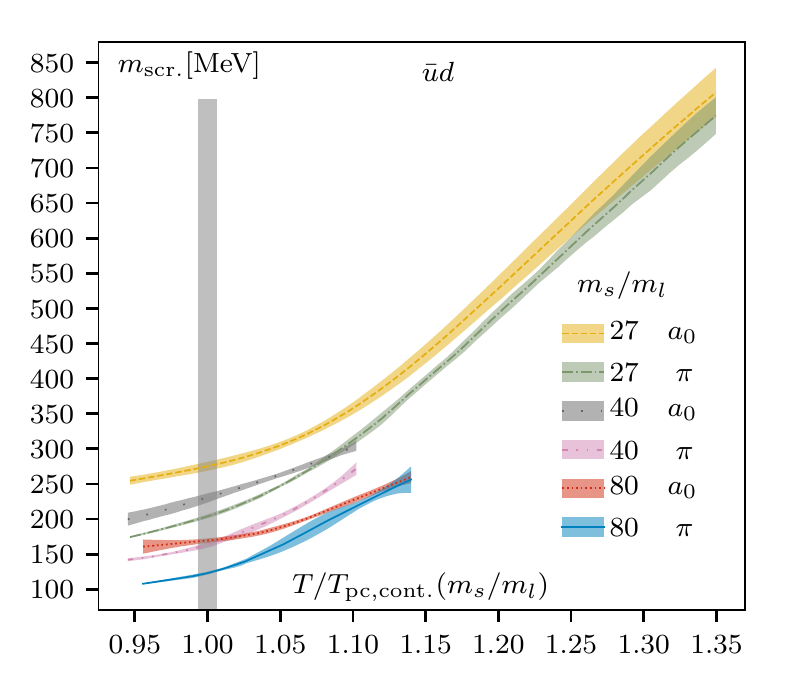}
\includegraphics[width=0.38\textwidth]{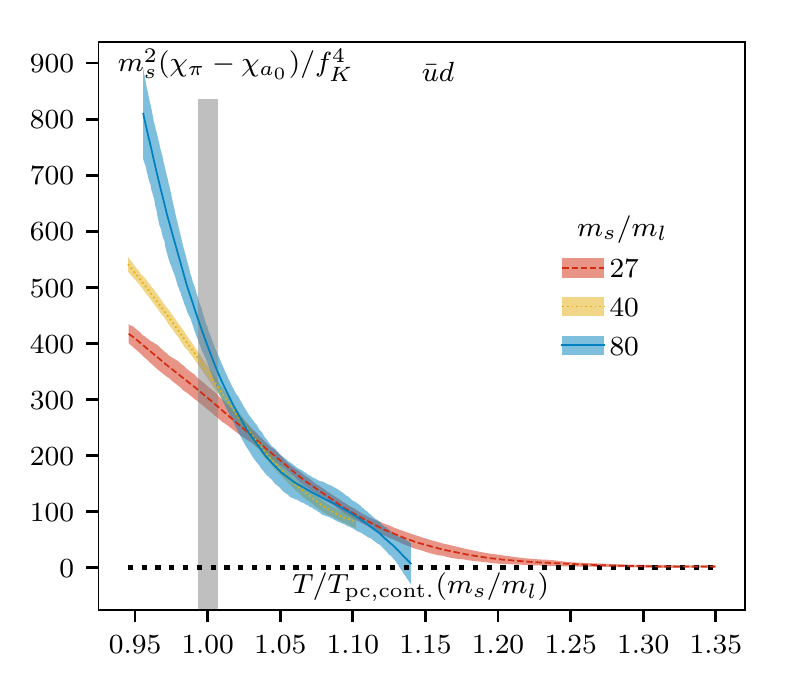} 
\caption{Screening masses (left) and 
the related susceptibilities (right)
of scalar and pseudo-scalar mesons
\cite{Bazavov:2019www,Dentinger:2021khg}.}
\label{fig:scrmass}
\end{figure}
More easily accessible are so-called
screening masses, which also are obtained from ordinary hadron correlation functions
and can be analyzed close to
the chiral limit.
Rather than analyzing the long-distance behavior of hadron
correlation functions
in Euclidean time, one extracts a so-called screening mass
from the long-distance behavior in one of the spatial
directions \cite{Detar:1987kae,Detar:1987hib}.
Finite temperature meson screening correlators, 
projected onto lowest
Matsubara frequency of a bosonic state, $p_0\equiv \omega_0=0$,
and zero transverse momentum, 
${\bf p}_\perp\equiv (p_x,p_y)=0$, are defined by
\begin{eqnarray}
G_\Gamma(z,T) &=& \int_0^\beta d\tau \int dxdy\,\Big\langle {\cal M}_\Gamma(\vec{r},\tau) \overline{{\cal M}_\Gamma}(\vec{0},0) \Big\rangle
\nonumber \\
&{\raise0.6ex\hbox{$\sim$\kern-0.75em\raise-1.4ex\hbox{\hspace*{-9pt}$z \to \infty$}}}&
        \hspace*{0.2cm}{\rm e}^{-m_\Gamma (T) z}\;\; , \;\; \vec{r}\equiv (x,y,z)\;\; ,
\label{eq:definition}
\end{eqnarray}
where ${\cal M}_\Gamma \equiv \bar{\psi} \Gamma \psi$ is a meson operator
that projects onto a quantum number channel that is selected through an appropriate choice of $\Gamma$-matrices
\cite{Bazavov:2019www,Detar:1987kae}. 
At large distances this permits the extraction of the screening mass, $m_\Gamma$, in the quantum number channel selected by $\Gamma$ 
from the exponential 
fall-off of these correlation functions.
In Fig.~\ref{fig:scrmass} (left) we show 
results for the scalar and pseudo-scalar screening
masses obtained in $(2+1)$-flavor QCD calculations 
for different values of the light to strange quark mass ratio. The integrated correlation functions define susceptibilities in these
quantum number channels, which also 
should be degenerate, if $U(1)_A$ is 
effectively restored. Both observables
seem to suggest that there remains
a significant remnant of $U(1)_A$ breaking at the chiral phase 
transition temperature, $T_c^0$, which
however reduces quickly above
the chiral transition and gives rise  to an effective restoration of $U(1)_A$ at $T\simeq 1.1 T_c^0$.

In the region $T>T_c^0$
the difference between pseudo-scalar and scalar susceptibilities is 
related to the so-called disconnected
part, $\chi_{dis}$, of the chiral susceptibility, $\chi_M = \chi_{dis} + \chi_{con} $, with
\begin{eqnarray}
    \chi_{dis} &=& \frac{1}{4N_\tau N_\sigma^3} \left( \langle ({\rm Tr} M_\ell^{-1})^2\rangle 
    -\langle {\rm Tr} M_\ell^{-1}\rangle^2 \right) \; ,
    \\
    \chi_{con} &=& \frac{1}{2N_\tau N_\sigma^3} \langle {\rm Tr} M_\ell^{-2}\rangle \; .
\end{eqnarray}

\begin{figure}[t]
\centering
\includegraphics[width=0.55\textwidth]{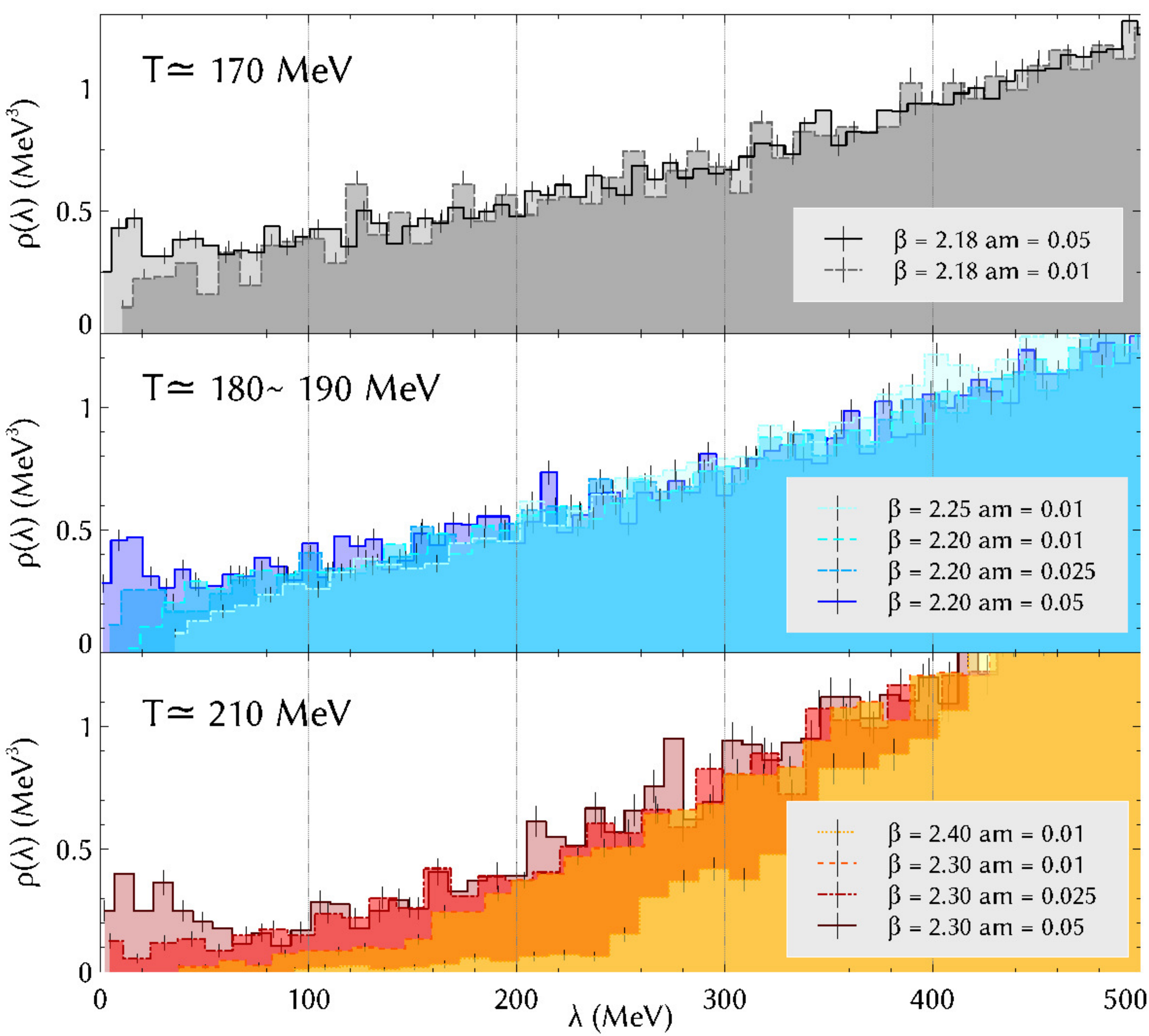}
\caption{Eigenvalue density of the overlap fermion matrix obtained in calculations with dynamical overlap fermions
\cite{Cossu:2013uua}.
}
\label{fig:over}
\end{figure}

The disconnected chiral susceptibility
can be expressed by an integral over
the eigenvalue density, $\rho(\lambda)$, of the fermion
matrix $M_f$,
\begin{equation}
  \chi_{dis} = \int_0^\infty d\lambda\,\rho(\lambda)
  \frac{2m_\ell^2}{(\lambda^2+m_\ell^2)^2} \; .
\label{chidis-eigen}
\end{equation}
In the chiral symmetric phase the density 
of vanishing eigenvalues, $\rho(0)$, 
vanishes.
In order for $\chi_{dis}$ to be nonetheless
non-zero in the chiral limit, the density
of near-zero eigenvalues needs to converge
to a non-vanish\-ing value ($\delta$-function) at $\lambda=0$ in the limit $m_\ell \rightarrow 0$ and $V\rightarrow \infty$. 
Controlling the various limits 
involved and also taking into 
account that the pseudo-critical transition temperature, $T_{pc}(H)$,
has a sizeable quark mass dependence
is difficult. Nonetheless,
studies of the temperature 
dependence of the eigenvalue density
of the Dirac matrix are crucial
for a detailed understanding of 
the influence of the
$U(1)_A$ anomaly on the QCD phase transition. Not surprisingly, it turns
out that at non-zero values of the
lattice spacing the spectrum of low lying eigenvalues is quite sensitive to the fermion discretization scheme.
Using fermions with good chirality even at non-zero lattice spacing
seems to be advantageous, although 
after having performed the extrapolation to the chiral limit, they should lead to results identical with those obtained, e.g. within the
staggered discretization scheme. 
Current results are ambiguous. 
We 
show in Fig.~\ref{fig:over} results
from a calculation of eigenvalue
distributions obtained from calculations with dynamical overlap
fermions \cite{Cossu:2013uua,Tomiya:2016jwr}. 
These calculations provide evidence
for a large density of near-zero eigenvalues and a non-zero eigenvalue density, possibly building up 
at $\lambda=0$. This is in contrast to
calculations performed with domain
wall fermions \cite{Buchoff:2013nra} as
well as so-called partially quenched calculations that use the overlap fermion
operator to calculate eigenvalue distributions on gauge field configurations generated with dynamical staggered fermions \cite{Dick:2015twa}.
Obviously this subtle aspect of the chiral
phase transition is not yet resolved and
the analysis of
$U(1)_A$ restoration will remain to be a central topic in finite temperature QCD in the years to come.

\section{The chiral phase transition at non-vanishing chemical potential}

In the studies of QCD at non-vanishing baryon chemical potential the 
search for the existence of a second order phase transition
at physical values of the quark masses, the critical end point (CEP),
finds particular attention. It separates the crossover regime at small
values of the chemical potential from a region of first order phase transitions,
which is predicted in many phenomenological models to exist at high density. 
The CEP is searched for extensively in heavy ion experiments and, if confirmed,
would provide a solid prediction for the existence of first order phase 
transitions in dense stellar matter, e.g. in neutron stars.

\begin{figure}[t]
\begin{center}
     \includegraphics[width=0.48\textwidth]{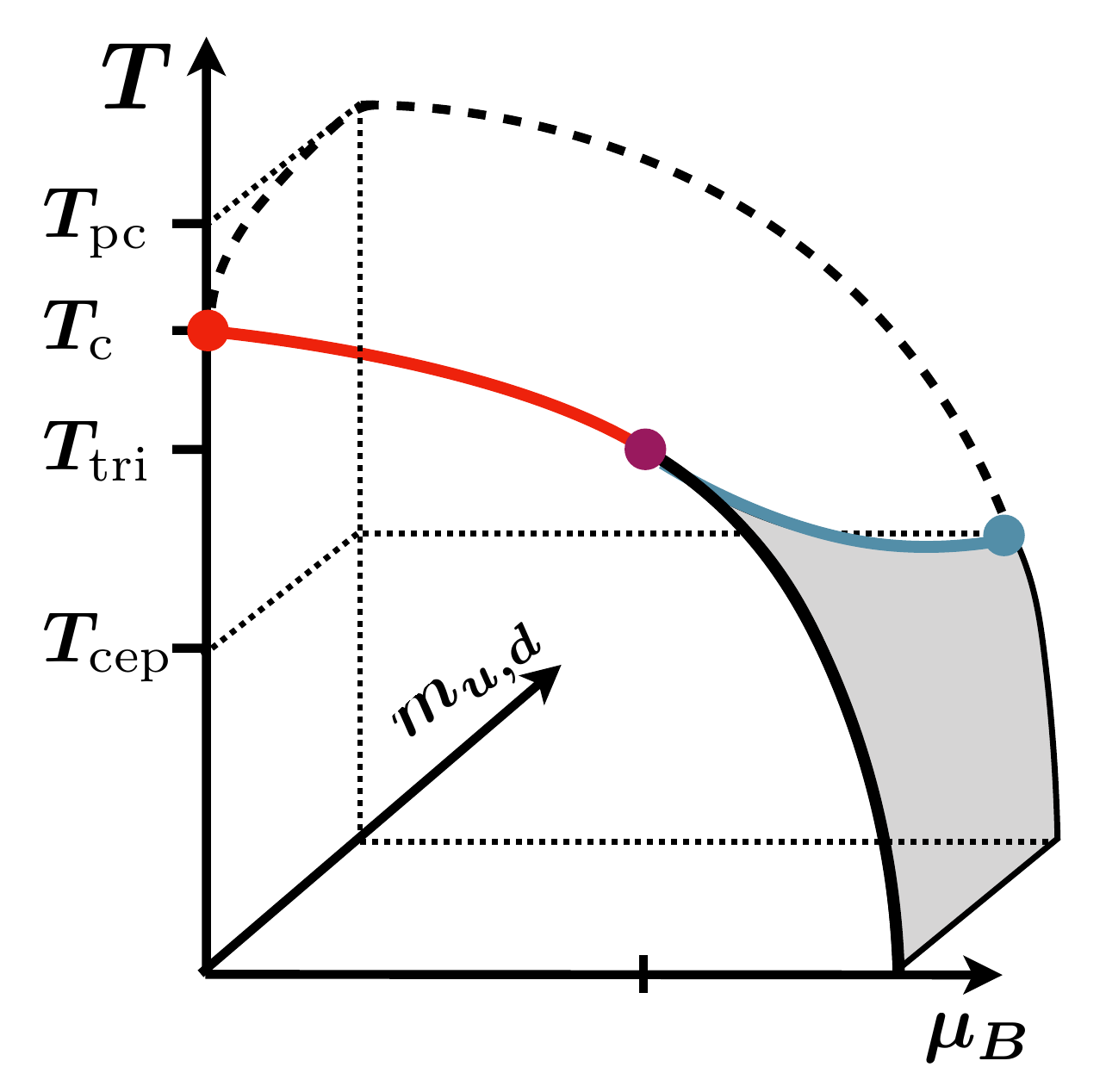}
        \end{center}
\caption{Sketch of a possible QCD phase diagram in the space of temperature ($T$), baryon chemical potential ($\mu_B$) and light quark masses ($m_{u,d}$).
        }
\label{fig:3dphase}
\end{figure}

The dependence of the transition temperature on the chemical
potentials, e.g. $T_{pc}(\mu_B)$, 
can be deduced from the $\mu_B$-dependent shift of the peak
in the chiral susceptibility.
At non-vanishing values of the 
baryon chemical potential, $\mu_B$,
the QCD phase transition temperature
in the chiral limit as well as the region of pseudo-critical behavior in 
QCD with its physical quark mass values shifts to smaller values of the temperature.  This
shift has been determined in calculations with imaginary values of 
the chemical potentials as well as 
from Taylor series expansions of the order parameter $M$ and its
susceptibility $\chi_M$.
Using a Taylor series ansatz for
$T_{pc}(\mu_B)$,
\begin{equation}
    T_c(\mu_B)= T_c^0\left( 1 -\kappa_2^B\left(\frac{\mu_B}{T_c^0}\right)^2
    -\kappa_4^B\left(\frac{\mu_B}{T_c^0}\right)^4
    \right)
\end{equation}
one finds for the curvature coefficients $\kappa_2^B\simeq 0.012$
while the next correction is consistent with zero in all current
studies, e.g. $\kappa_4^B= 0.00032(67)$ \cite{Borsanyi:2020fev}. The 
pseudo-critical temperature $T_{pc}$
at physical values of the light and strange quark masses thus drops to
about 150~MeV at $\mu_B\simeq 2 T_{pc}$.
This is still considerably larger than
the chiral phase transition temperature, $T_c^0$, determined at $\mu_B=0$.
As various model calculations
\cite{Halasz:1998qr,Buballa:2018hux}
suggest that the CEP at non-zero $\mu_B$ is located at a temperature below $T_c^0$ one thus needs
to get access to thermodynamics at large chemical potentials. Assuming
that the curvature of the 
pseudo-critical line does not change 
drastically at large values of the
chemical potentials, our current
understanding of the QCD phase diagram in the $m_\ell$-$T$-$\mu_B$
space (see Fig.~\ref{fig:3dphase})
suggests that a possible CEP 
in the phase diagram may exist 
only at a temperature,
\begin{equation}
 T^{CEP}(\mu_B^{CEP}) < 130~{\rm MeV}
 \; , \;
 \mu_B^{CEP} > 400~{\rm MeV}\; .
\end{equation}
Reaching the region $\mu_B/T > 3$
is a major challenge for any of the
currently used approaches in lattice
QCD calculations as well as for 
collider based 
heavy ion experiments that search
for the CEP.

\section{Equation of state of strongly interacting matter}

The equation of state (EoS) of strongly interacting matter, {\it i.e.} the pressure and 
its derivatives with respect to temperature and chemical 
potentials provides the basic 
information on the phase structure of
QCD. It is of central importance 
not only for the analysis of critical
behavior in QCD but also for the
analysis of experimental results 
on strong interaction thermodynamics 
that are obtained in relativistic
heavy ion collision experiments.

\begin{figure*}[t]
\centering
\includegraphics[width=0.42\textwidth]{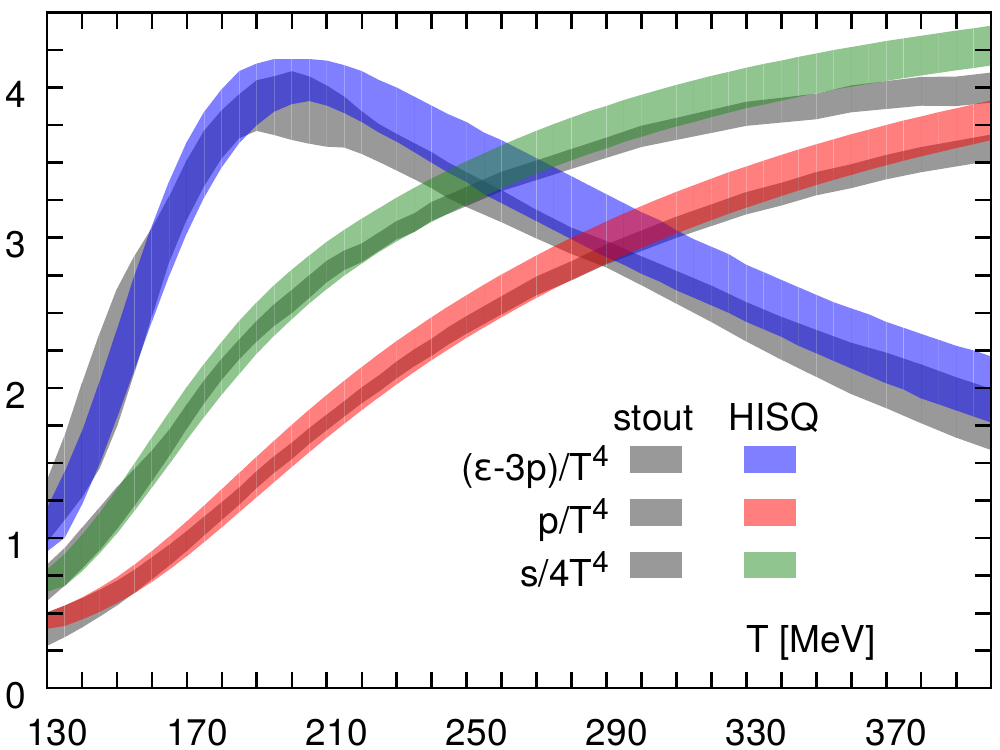}\hspace{0.8cm}
\includegraphics[width=0.42\textwidth]{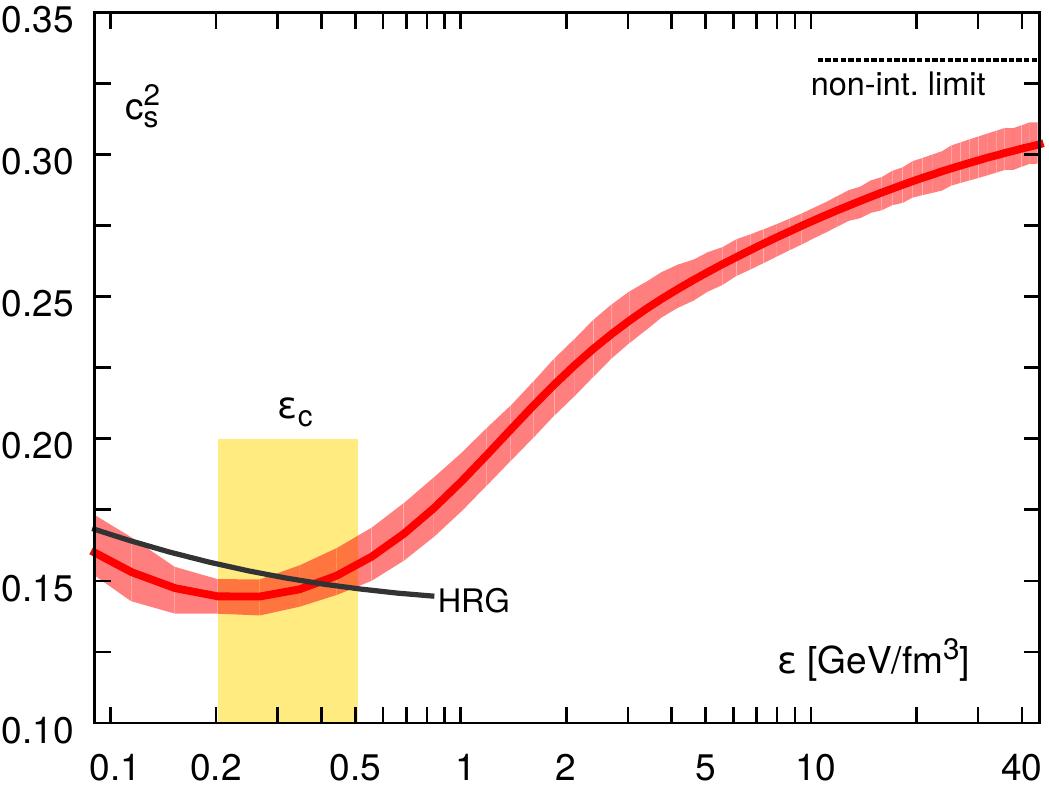}
\caption{{\it Left:} Pressure, energy and entropy densities in (2+1)-flavor QCD at vanishing 
chemical potential. The figure is taken
from \cite{HotQCD:2014kol}. Also shown in the figure are results obtained with the stout discretization scheme for staggered 
fermions \cite{Borsanyi:2013bia}.
{\it Right:} The speed of sound as 
function of energy density.
}
\label{fig:EoS}
\end{figure*}

At vanishing values of the chemical
potentials the QCD EoS
is well controlled and consistent results for pressure, energy and entropy densities, as well as derived observables such as the speed of sound
or specific heat,
have been obtained by several groups
\cite{Borsanyi:2013bia,HotQCD:2014kol}.
We show results for some of these observables in Fig.~\ref{fig:EoS}. The figure
on the right shows the square of  the speed of sound, $c_s^2$, as function of the 
energy density. It can be seen that
$c_s^2$ has a minimum in the transition region, sometimes called the softest point of the QCD EoS
\cite{Hung:1994eq}. The energy density in the vicinity of the pseudo-critical temperature ($T_{pc} \simeq 155$~MeV)
is found to be,
\begin{equation}
    \epsilon_{\rm c} \simeq (350\pm 150)~{\rm MeV}/{\rm fm^3}
    \; ,
\end{equation}
which is compatible with the energy
density of the nucleon, $m_N/(4\pi r_N^3/3)$ for nucleon radii in the range $r_N= (0.8-1)$~fm. Also shown
in the top figure is the trace of 
the energy-momentum tensor, $(\epsilon-3P)/T^4$. Its deviation 
from zero gives some hint to the 
relevance of interactions in the medium (for an ideal gas as well as to leading order in high temperature perturbation theory one has $\epsilon=3P$). Not unexpected this is largest close to the transition region and decreases only slowly 
in the high temperature regime. This
large deviations from ideal gas or perturbative behavior is seen in many
observables at temperature $T_{pc}< T<2 T_{pc}$.

\begin{figure*}[t]
\centering
\includegraphics[width=0.52\textwidth]{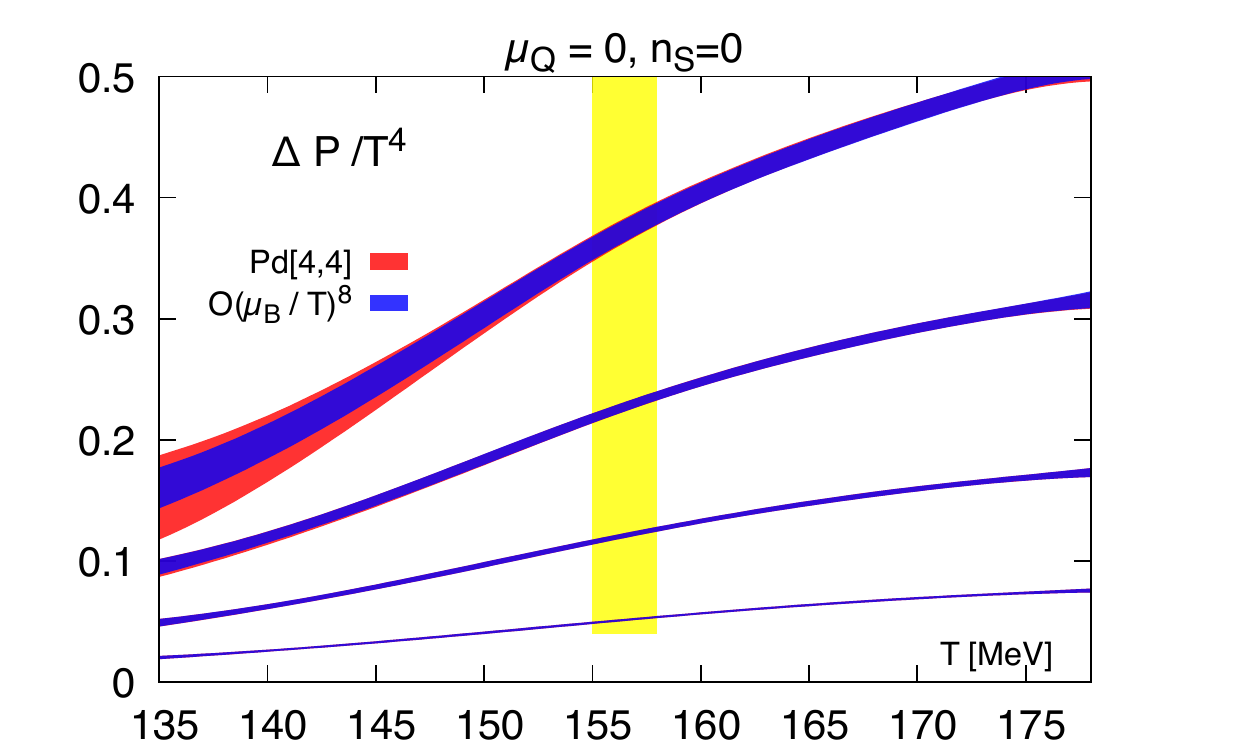}\hspace*{-0.5cm}
\includegraphics[width=0.52\textwidth]{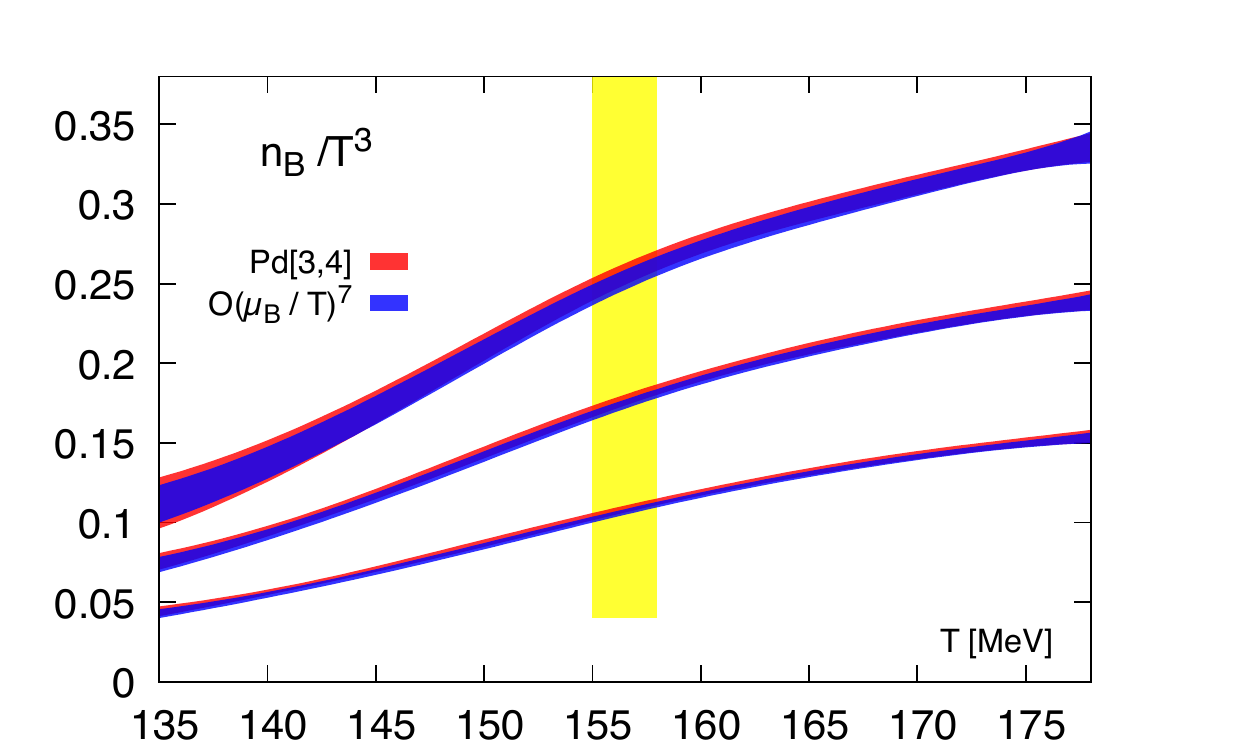}
\caption{$\mu_B$-dependent contribution to the pressure (left) and net baryon number density (right) in (2+1)-flavor QCD at several values of the baryon chemical potential
chemical potential, $\mu/T_B=1.0$, 1.5, 2.0, 2.5, (bottom to top) and
for $\hat{\mu}_B=2.0$.
Shown are results from  Taylor expansion up to eighth order in $\hat{\mu}_B$ in the pressure series
for isospin symmetric ($\mu_Q=0$)
strangeness neutral ($n_S=0$) matter
and corresponding Pad\'e
approximants obtained from these Taylor expansion coefficients. The figures are taken
from \cite{Bollweg:2022rps}.
}
\label{fig:EoSmu}
\end{figure*}

Calculations of the equation of 
state as a function of $T$ and $\mu_B$
have been performed using direct simulations at imaginary chemical potentials, which then get analytically continued to real values
of the chemical potentials \cite{Borsanyi:2012cr}, as well
as calculations using up to eighth order Taylor expansions in $\mu_B$ \cite{Bollweg:2022rps}.
Results of such calculations agree 
well for $\mu_B/T\le (2-2.5)$.
In Fig.~\ref{fig:EoSmu} we show
results for the $\mu_B$-dependent contribution to the pressure and 
net baryon number density. Comparing 
Fig.~\ref{fig:EoSmu}~(left) with 
Fig.~\ref{fig:EoS}~(left) shows that
at $\mu_B/T\simeq 2$ and $T\simeq T_{pc}$ the pressure increases by 
about 30\%, which is due to 
the increase in number of baryons in 
the medium.

At larger values of the baryon chemical potential the Taylor series
will not convergence due
to the presence of either poles in the complex $\mu_B$-plane or a real 
pole, that may correspond to the searched for CEP. The occurrence of 
poles in the complex plane also generates problems for the analytic
continuation of results obtained in 
simulations at imaginary values of $\mu_B$ as a suitable ansatz for the 
continuation needs to be found.
Many approaches to improve over
straightforward Taylor series approaches or simulations at imaginary chemical potential are
currently being discussed 
\cite{Mukherjee:2019eou,Mondal:2021jxk,Dimopoulos:2021vrk,Borsanyi:2022qlh}.
In the context of Taylor expansions 
a natural way to proceed is to use 
Pad\'e approximants, which provide
a resummation of the Taylor series and 
reproduce this series, when expanded for small $\mu_B$ \cite{Datta:2016ukp,Bollweg:2022rps}. Results from 
 [4,4] and [3,4] Pad\'e approximants for the
pressure and number density series, respectively, are also shown in Fig.~\ref{fig:EoSmu}. 
The good agreement with the Taylor series 
for $\mu_B/T\le 2.5$ gives confidence in the validity of the Taylor series
results and once more seems to rule out the occurrence of a CEP in this
parameter range.

\section{Outlook}
Achieving better control over 
the influence of the axial anomaly on the QCD phase transition in the chiral limit at vanishing chemical potentials and getting better control
over the structure of the QCD phase diagram at large non-zero values of the chemical potentials certainly are the two largest challenges in studies of QCD thermodynamics for the next decade.

\section*{Acknowledgements}
This work was supported by
the DFG Collaborative Research Centre 315477589-TRR 211, ”Strong interaction matter under extreme
conditions”.

\bibliography{Karsch}
\end{document}